\documentclass[journal]{IEEEtran}
\IEEEoverridecommandlockouts
\usepackage{cite}
\usepackage{amsmath,amsthm,amssymb,amsfonts}
\usepackage{array}
\usepackage{graphicx}
\usepackage{textcomp}
\usepackage[mathscr]{eucal}
\usepackage[table]{xcolor}
\usepackage{xcolor}
\usepackage{cleveref}
\crefformat{section}{\S#2#1#3}
\usepackage[framemethod=tikz]{mdframed}
\usepackage[ruled,vlined,linesnumbered]{algorithm2e}
\usepackage{algpseudocode}
\usepackage{multirow}
\usepackage{rotating}
\newtheorem{definition}{Definition}
\usepackage[font=small]{caption}
\usepackage{pifont}
\usepackage[flushmargin]{footmisc}

\algnewcommand\LeftComment[2]{%
\hspace{#1\algindent}$\triangleright$ \eqparbox{COMMENT}{#2} \hfill %
}
\let\oldnl\nl
\newcommand{\nonl}{\renewcommand{\nl}{\let\nl\oldnl}}

\def\BibTeX{{\rm B\kern-.05em{\sc i\kern-.025em b}\kern-.08em
    T\kern-.1667em\lower.7ex\hbox{E}\kern-.\section{Conclusion}125emX}}
\begin{document}

\title{Analysis, Modeling, and Representation  of COVID-19 Spread: A Case Study on India}
\author{\IEEEauthorblockN{Rahul Mishra, Hari Prabhat Gupta, and Tanima Dutta} \vspace{-0.5cm}
 \thanks{The authors are with the Department of Computer Science and Engineering, Indian Institute of Technology (BHU) Varanasi, India (e-mail: rahulmishra.rs.cse17@iitbhu.ac.in;  hariprabhat.cse@iitbhu.ac.in; tanima.cse@iitbhu.ac.in)}}
\maketitle
\begin{abstract} 
Coronavirus outbreak is one of the most challenging pandemics for the entire human population of the planet Earth. Techniques such as the isolation of infected persons and maintaining social distancing are the only preventive measures against the epidemic COVID-19. The actual estimation of the number of infected persons with limited data is an indeterminate problem faced by data scientists. There are a large number of techniques in the existing literature, including reproduction number, the case fatality rate, \text{etc.}, for predicting the duration of an epidemic and infectious population. This paper presents a case study of different techniques for analysing, modelling, and representation of data associated with an epidemic such as COVID-19. We further propose an algorithm for estimating infection transmission states in a particular area. This work also presents an algorithm for estimating end-time of an epidemic from Susceptible Infectious and Recovered model. Finally, this paper presents empirical and data analysis to study the impact of transmission probability, rate of contact, infectious, and susceptible on the epidemic spread.
\end{abstract}

\begin{IEEEkeywords}
 Coronavirus, epidemic, infectious, isolation, susceptible.
\end{IEEEkeywords}

\section{Introduction}
The disease is an abnormal condition that has stricken human civilisation since their existence~\cite{de2020impact}. Every disease is the medical condition that has some symptoms and signs for their identification. These symptoms and signs provide the suitability to identify the disease in the early stage. However, some diseases do not have identifiable symptoms, which makes them severe. Due to significant growth in medical science, most of the existing diseases are cured or halted from further damage to the human body. However, some diseases spread due to some unknown virus or bacteria and lead to the most deadly pandemic in humans. One of such pandemic is novel coronavirus named COVID-19 that has potential for maximal death~\cite{rezaeetalab2020covid, 9043580}. The controlling mechanism to hamper the spread of an epidemic like COVID-19 requires an attentive and strict application of quarantine and social distancing. The effect of the epidemic (coronavirus) spread, quarantine, and social distancing can be simulated using data analysis, modelling, and representation techniques of data mining~\cite{4358925,giordano2020modelling,anastassopoulou2020data}. These techniques help in predicting disease behaviour and its corresponding preventive measure. 

One of the most widely used techniques for predicting the behaviour of an epidemic in term of infection spread during an epidemic is the reproduction number ($R_o$). It estimates the spread of secondary infections from the primary infected persons. Next, the case fatality rate captures the number of deaths due to an epidemic spread. The case fatality rate plays a vital role in estimating the actual damage due to an epidemic. In other words, the number of people that are infected (estimated through reproduction number) losses their life during an epidemic. Further, there are four states of infection spread in an epidemic, \textit{i.e.,} no contact, local, untraceable, and source missing transmission. The first state is the least harmful and often occur at the initial stage of the spread, whereas, the last state is the critical transmission state indicating the spread of infection is severe. In the literature for assessment of epidemic spread, different models are discussed like Susceptible Infectious Recovered (SIR)~\cite{8957067}, Susceptible Infectious (SI), and Susceptible Infectious and Susceptible (SIS) model. 

\begin{figure}[h]
\centering
\includegraphics[scale=1]{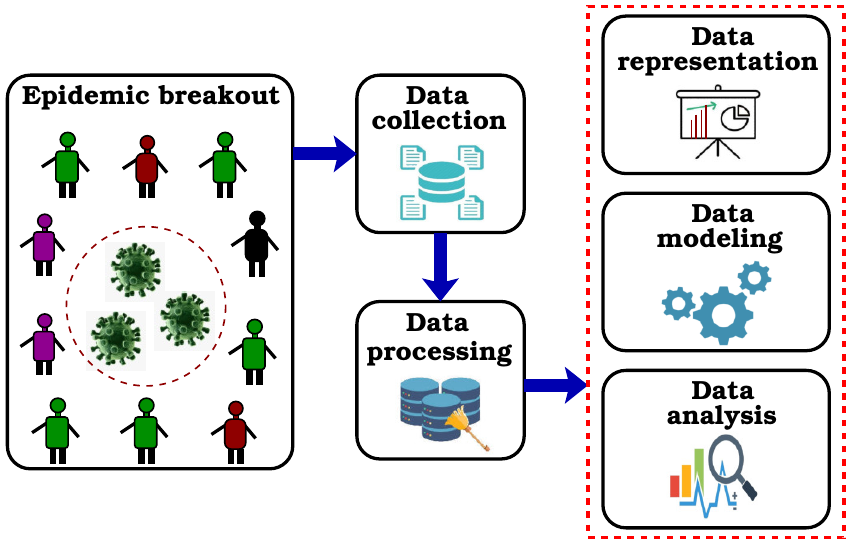}
\caption{Illustration of data analysis, modelling and representation with the prior step of data collection and processing.}
\label{master1}
\end{figure}

The modern era of information encourages us to perform data analysis for predicting the future course of actions against disease and its behaviour~\cite{5416718}. The prediction is performed by a model that takes the historical data to learn a pattern (like the pattern of epidemic spread COVID-19). The performance of these models depends heavily upon data representation and storage. Therefore, in this paper, we have summarises different techniques presented in the literature for the estimation of various factors that are responsible for an epidemic (COVID-19).  We further discuss the methods that are predicting the end time of an epidemic and the total population that is going to be infected and susceptible. This is the first work as a case study that covers data analysis, modelling and representation of the coronavirus spread in India. Fig.~\ref{master1} illustrate the steps involved prior to the data analysis, modelling, and representation. After collecting the epidemic data, the first and foremost task is to perform data processing (like data cleaning, normalization, \textit{etc}). Further, analysis and modelling is performed on the processed data. The papers answer the following question for predicting the spread of infection and estimating the duration of an epidemic. 

\begin{itemize}
\item\textbf{Question 1:} What is the role of reproduction number in the estimation of the population infected during an epidemic such as COVID-19? How to determine the average of the reproduction number for an epidemic disease dataset? (Section~\ref{repro})
\item\textbf{Question 2:} Is the reproduction number capable of estimating the number of infected people in a particular area during an epidemic? What is the impact of the case fatality rate on estimating the damage due to an epidemic like COVID-19? (Section~\ref{fatality})
\item\textbf{Question 3:} What are the infection transmission states in an epidemic? How to identify the transmission state in a particular area? Is there any existing algorithm for transmission state identification? (Section~\ref{state1})
\item\textbf{Question 4:} How to model an epidemic for accurate estimation of infection spread and recovery rate?  Which model is suitable for end-time estimation of an epidemic? What is the Susceptible Infectious Recovered model? How  Susceptible Infectious Recovered (SIR), Susceptible Infectious (SI), and Susceptible Infectious and Susceptible (SIS) models estimate the impact of an epidemic? (Section~\ref{model1})
\end{itemize}

The rest of the paper is organised as follows. Next section illustrates Covid-19 open access dataset, the parameters used for measuring the impact of an epidemic, and the method to identify the state of the epidemic. Section~\ref{model1} illustrates the mathematical models of the epidemic. Section~\ref{res1} illustrates the empirical results and paper is concluded in Section~\ref{con1}. 

\section{Dataset and Parameters for measuring the spread of an epidemic disease}\label{measurment}
\begin{figure}[h]
\centering
\includegraphics[scale=.35]{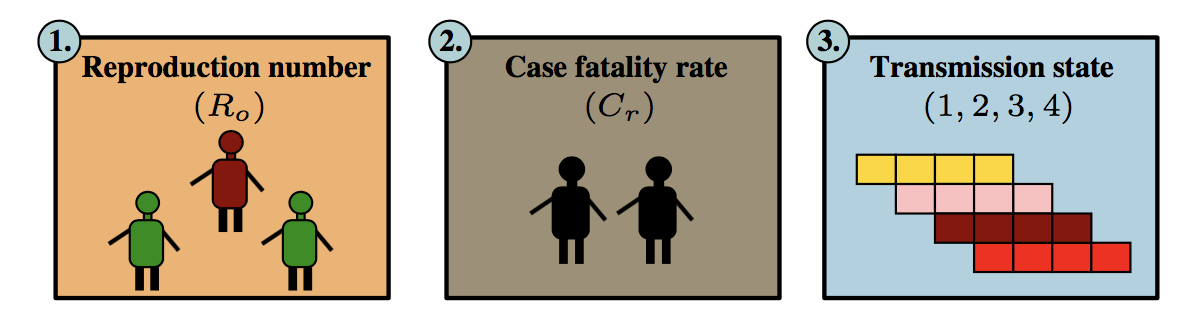}
\caption{Illustration of parameters for measuring the epidemic.}
\label{param1}
\end{figure}

In this section, we first illustrate an open access dataset of Covid-19 that we use in the rest of the paper. We next define the parameters used for measuring the impact of epidemic, as shown in Fig.~\ref{param1}.

\subsection{Dataset of Covid-19}
This work uses COVID-$19$ dataset available at~\cite{data} for predicting and analyzing the outbreak of epidemic in India. 
The dataset illustrates the detail description of the patients infected from coronavirus. Table~\ref{data_table} depicts the first $10$ entries of the dataset having different columns including, patient number, state patient number, date announcement, age bracket, gender, detected city, detected district, detected state, state code, current status, notes, contacted from which patient, nationality, transmission type, and status change date. The column header \textit{patient number} indicates the actual patient count in India, and \textit{state patient number} is a combination of state code, detected district code and its patient count. For example, the infected patient $1$ is from Thrissur district (district code: TS) of Kerala (state code: KL). Therefore, state patient number is KL-TS-P1 for patient $1$. The column \textit{current status} indicates the current state of the patient, \textit{i.e.,}  recovered, hospitalized, or deceased. The column with header \textit{notes} holds the details about travelling history of the patient. Next, the local transmission from one patient to another is recorded in column \textit{ contacted from which patient}. Further, \textit{nationality} followed by \textit{transmission type} that depicts whether the transmission of infection is imported from other countries or local. Finally, the column \textit{status change date} illustrates the status change of infected patient to recovered or deceased from infected.

 \begin{table*}[h]
\centering
\caption{Dataset of Covid-19~\cite{data} (total rows are 20000+ as on April 23, 2020).}
\resizebox{1.000\textwidth}{!}{
\begin{tabular}{|c|c|c|c|c|c|c|c|c|c|c|c|c|c|c|}
\hline
\textbf{\begin{tabular}[c]{@{}c@{}}Patient \\ Number\end{tabular}} & \textbf{\begin{tabular}[c]{@{}c@{}}State \\ Patient \\ Number\end{tabular}} & \textbf{\begin{tabular}[c]{@{}c@{}}Date \\ Announced\end{tabular}} & \textbf{\begin{tabular}[c]{@{}c@{}}Age \\ Bracket\end{tabular}} & \textbf{Gender} & \textbf{\begin{tabular}[c]{@{}c@{}}Detected \\ City\end{tabular}} & \textbf{\begin{tabular}[c]{@{}c@{}}Detected \\ District\end{tabular}} & \textbf{\begin{tabular}[c]{@{}c@{}}Detected \\ State\end{tabular}} & \textbf{\begin{tabular}[c]{@{}c@{}}State \\ code\end{tabular}} & \textbf{\begin{tabular}[c]{@{}c@{}}Current \\ Status\end{tabular}}  & \textbf{Notes}                & \textbf{\begin{tabular}[c]{@{}c@{}}Contracted from \\ which Patient \\ (Suspected)\end{tabular}} & \textbf{Nationality} & \textbf{\begin{tabular}[c]{@{}c@{}}Type of \\ transmission\end{tabular}} & \textbf{\begin{tabular}[c]{@{}c@{}}Status \\ Change \\ Date\end{tabular}} \\ \hline
1                                                                  & KL-TS-P1                                                                          & 30/01/2020                                                            & 20                                                                 & F               & Thrissur                                                             & Thrissur                                                                 & Kerala                                                                & KL                                                                & Recovered               & Travelled from Wuhan          &                                                                                                        & India                & Imported                                                                    & 14/02/2020                                                                      \\ \hline
2                                                                  & KL-AL-P1                                                                          & 02/02/2020                                                            &                                                                    &                 & Alappuzha                                                            & Alappuzha                                                                & Kerala                                                                & KL                                                                & Recovered               & Travelled from Wuhan          &                                                                                                        & India                & Imported                                                                    & 14/02/2020                                                                      \\ \hline
3                                                                  & KL-KS-P1                                                                          & 03/02/2020                                                            &                                                                    &                 & Kasaragod                                                            & Kasaragod                                                                & Kerala                                                                & KL                                                                & Recovered               & Travelled from Wuhan          &                                                                                                        & India                & Imported                                                                    & 14/02/2020                                                                      \\ \hline
4                                                                  & DL-P1                                                                             & 02/03/2020                                                            & 45                                                                 & M               & East Delhi                                                           & East Delhi                                                               & Delhi                                                                 & DL                                                                & Recovered               & Travelled from Austria &                                                                                                        & India                & Imported                                                                    & 15/03/2020                                                                      \\ \hline
5                                                                  & TS-P1                                                                             & 02/03/2020                                                            & 24                                                                 & M               & Hyderabad                                                            & Hyderabad                                                                & Telangana                                                             & TG                                                                & Recovered               & Travelled from Dubai          &                                                                                                        & India                & Imported                                                                    & 02/03/2020                                                                      \\ \hline
6                                                                  &                                                                                   & 03/03/2020                                                            & 69                                                                 & M               & Jaipur                                                               & Jaipur                                                                   & Rajasthan                                                             & RJ                                                                & Recovered               & Travelled from Italy          &                                                                                                        & Italy                & Imported                                                                    & 03/03/2020                                                                      \\ \hline
7                                                                  &                                                                                   & 04/03/2020                                                            & 55                                                                 &                 & Gurugram                                                             & Gurugram                                                                 & Haryana                                                               & HR                                                                & Recovered               & Travelled from Italy          & P6                                                                                                     & Italy                & Imported                                                                    & 29/03/2020                                                                      \\ \hline
8                                                                  &                                                                                   & 04/03/2020                                                            & 55                                                                 &                 & Gurugram                                                             & Gurugram                                                                 & Haryana                                                               & HR                                                                & Recovered               & Travelled from Italy          & P6                                                                                                     & Italy                & Imported                                                                    & 29/03/2020                                                                      \\ \hline
9                                                                  &                                                                                   & 04/03/2020                                                            & 55                                                                 &                 & Gurugram                                                             & Gurugram                                                                 & Haryana                                                               & HR                                                                & Recovered               & Travelled from Italy          & P6                                                                                                     & Italy                & Imported                                                                    & 29/03/2020                                                                      \\ \hline
10                                                                 &                                                                                   & 04/03/2020                                                            & 55                                                                 &                 & Gurugram                                                             & Gurugram                                                                 & Haryana                                                               & HR                                                                & Recovered               & Travelled from Italy          & P6                                                                                                     & Italy                & Imported                                                                    & 29/03/2020                                                                      \\ \hline
\end{tabular}
}
\label{data_table}
\end{table*}

\subsection{Epidemic first parameter: Reproduction number}\label{repro}
\begin{figure}[h]
\centering
\includegraphics[scale=1]{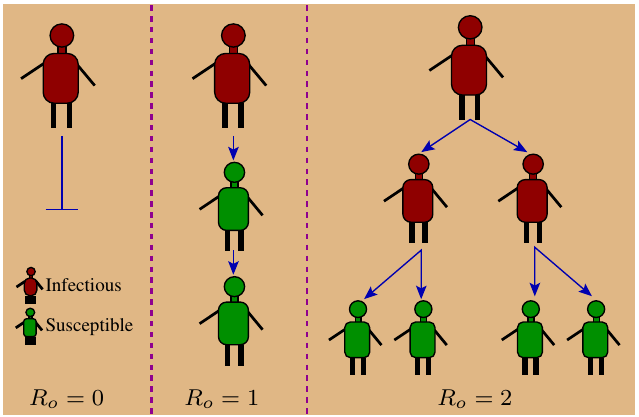}
\caption{An example scenario of basic reproduction number $R_o$.}
\label{example}
\end{figure}

The epidemic disease modelling incorporates the reproduction number, denoted by $R_o$, to understand the disease outbreak by wrapping up all the infection transmission in a specified area. The reproduction number specifies the number of persons being infected by an infectious person (a person suffering from the disease)~\cite{zhou2020preliminary}. It defines as
\begin{definition}
Reproduction number of a given region is the ratio of the total number of persons infected to the total number of persons who are infecting, \textit{i.e.},
\begin{equation}
R_o = \frac{\text{Total persons infected}}{\text{Total who are infecting}}.
\end{equation}
\end{definition}
Fig.~\ref{example} illustrates an example scenario where from left to right there is an increment in the value of $R_o$. \textcolor{black}{The number of persons infected by an infectious person} also increase in a similar pattern. The value of $R_o$ less then unity indicates that the epidemic elimination and greater than unity leading to an epidemic outbreak. 
\\

\begin{figure}[h]
\centering
\includegraphics[scale=1]{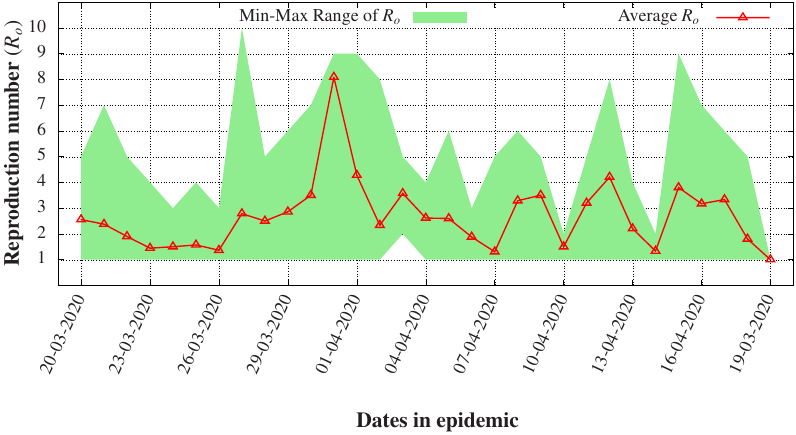}
\caption{Illustration of $R_o$ with time.}
\label{rograph}
\end{figure}

$\bullet$ \textbf{Importance of $R_o$}: $R_o$ is an essential metric that indicates how contiguous the disease is?. A disease is said to be highly contiguous when an infected individual transmits the infection to an unexpectedly large number of people. The mechanism of providing a proper treatment or appropriate handling of any disease requires full information about how the infection develops in the body,  the mode of transmission, and the social structure where it can spread. If all this information is available, then the outbreak can be detected and monitored quickly and allows us to react accordingly. The reproduction number $R_o$ captures all these necessary information by incorporating three components: population density, rate of infection, and rate of recovery or death~\cite{van2002reproduction}. 
\\
$\bullet$ \textbf{$R_o$ in given dataset}:  Fig.~\ref{rograph} illustrates the value of $R_o$ for the given day. It also shows the minimum and maximum $R_o$ in the states of \textcolor{black}{India on each day. We conclude from the result that the difference between the minimum and maximum value of $R_o$ is very high, which indicates that the epidemic is very high in some state while others are free. The result also illustrates that minimum value of $R_o$ increase with time which indicates that the epidemic has spread in India. Table~\ref{tab2} illustrates the value of $R_o$ for the spread of COVID-19 in different states of India from the dataset available at~\cite{data} up till April, $18$ $2020$. The table shows that few states like Haryana and Punjab have person to person spread of coronavirus reaches to $20$.} The average $R_o$ value of some state is less than $1$, which is a good indication of the lower spread of COVID-19 in India and sub-continents. However, some states show a value of $R_o$ more than $3$ that requires a significant effort in the spread control methodology in the country. The overall $R_o$ for India is $1.79$ that indicates the COVID-19 is in its initial stage in the country.

\begin{table*}[htb!]
\centering
\caption{State-wise reproductive number ($R_o$) using Covid-19 dataset~\cite{data}.}
\resizebox{.640\textwidth}{!}{
\begin{tabular}{|c|c|c|c|c|c|c|c|c|c|c|c|c|c|c|c|c|c|c|c|c|c|}
\hline
\multicolumn{22}{|c|}{\textbf{Andaman and Nicobar}}                                                                              \\ \hline
\textit{Person contact} & 1 & 2 & 3 & 4 & 5 & 6 & 7 & 8 & 9 & 10 & 11 & 12 & 13 & 14 & 15 & 16 & 17 & 18 & 19 & 20 & Avg.  \\ \hline
\textit{Count}          & 0 & 0 & 1 & 0 & 1 & 0 & 0 & 0 & 0 & 0  & 0  & 0  & 0  & 0  & 0  & 0  & 0  & 0  & 0  & 0  &  4     \\ \hline
\hline
\multicolumn{22}{|c|}{\textbf{Andhra Pradesh}}                                                                              \\ \hline
\textit{Person contact} & 1 & 2 & 3 & 4 & 5 & 6 & 7 & 8 & 9 & 10 & 11 & 12 & 13 & 14 & 15 & 16 & 17 & 18 & 19 & 20 & Avg.  \\ \hline
\textit{Count}          & 3 & 0 & 2 & 0 & 0 & 0 & 0 & 0 & 0 & 0  & 0  & 0  & 0  & 0  & 0  & 0  & 0  & 0  & 0  & 0  &  1.8   \\ \hline
\hline
\multicolumn{22}{|c|}{\textbf{Arunachal Pradesh}}                                                                              \\ \hline
\textit{Person contact} & 1 & 2 & 3 & 4 & 5 & 6 & 7 & 8 & 9 & 10 & 11 & 12 & 13 & 14 & 15 & 16 & 17 & 18 & 19 & 20 & Avg.  \\ \hline
\textit{Count}          & 0 & 0 & 0 & 0 & 0 & 0 & 0 & 0 & 0 & 0  & 0  & 0  & 0  & 0  & 0  & 0  & 0  & 0  & 0  & 0  &  0    \\ \hline
\hline
\multicolumn{22}{|c|}{\textbf{Assam}}                                                                              \\ \hline
\textit{Person contact} & 1 & 2 & 3 & 4 & 5 & 6 & 7 & 8 & 9 & 10 & 11 & 12 & 13 & 14 & 15 & 16 & 17 & 18 & 19 & 20 & Avg. \\ \hline
\textit{Count}          & 0 & 0 & 0 & 0 & 0 & 0 & 0 & 0 & 0 & 0  & 0  & 0  & 0  & 0  & 0  & 0  & 0  & 0  & 0  & 0  &  0   \\ \hline
\hline
\multicolumn{22}{|c|}{\textbf{Bihar}}                                                                              \\ \hline
\textit{Person contact} & 1 & 2 & 3 & 4 & 5 & 6 & 7 & 8 & 9 & 10 & 11 & 12 & 13 & 14 & 15 & 16 & 17 & 18 & 19 & 20 & Avg. \\ \hline
\textit{Count}          & 0 & 1 & 0 & 0 & 0 & 1 & 0 & 1 & 0 & 0  & 0  & 0  & 0  & 0  & 0  & 0  & 0  & 0  & 0  & 0  &  5.33    \\ \hline
\hline
\multicolumn{22}{|c|}{\textbf{Chandigarh}}                                                                              \\ \hline
\textit{Person contact} & 1 & 2 & 3 & 4 & 5 & 6 & 7 & 8 & 9 & 10 & 11 & 12 & 13 & 14 & 15 & 16 & 17 & 18 & 19 & 20 & Avg. \\ \hline
\textit{Count}          & 1 & 0 & 0 & 1 & 0 & 0 & 0 & 0 & 0 & 0  & 0  & 0  & 0  & 0  & 0  & 0  & 0  & 0  & 0  & 0  &  2.5   \\ \hline
\hline
\multicolumn{22}{|c|}{\textbf{Chattisgarh}}                                                                              \\ \hline
\textit{Person contact} & 1 & 2 & 3 & 4 & 5 & 6 & 7 & 8 & 9 & 10 & 11 & 12 & 13 & 14 & 15 & 16 & 17 & 18 & 19 & 20 & Avg. \\ \hline
\textit{Count}          & 0 & 0 & 0 & 0 & 0 & 0 & 0 & 0 & 0 & 0  & 0  & 0  & 0  & 0  & 0  & 0  & 0  & 0  & 0  & 0  &  0   \\ \hline
\hline
\multicolumn{22}{|c|}{\textbf{Dadra Nagar Haveli}}                                                                              \\ \hline
\textit{Person contact} & 1 & 2 & 3 & 4 & 5 & 6 & 7 & 8 & 9 & 10 & 11 & 12 & 13 & 14 & 15 & 16 & 17 & 18 & 19 & 20 & Avg. \\ \hline
\textit{Count}          & 0 & 0 & 0 & 0 & 0 & 0 & 0 & 0 & 0 & 0  & 0  & 0  & 0  & 0  & 0  & 0  & 0  & 0  & 0  & 0  &  0   \\ \hline
\hline
\multicolumn{22}{|c|}{\textbf{Delhi}}                                                                              \\ \hline
\textit{Person contact} & 1 & 2 & 3 & 4 & 5 & 6 & 7 & 8 & 9 & 10 & 11 & 12 & 13 & 14 & 15 & 16 & 17 & 18 & 19 & 20 & Avg. \\ \hline
\textit{Count}          & 2 & 0 & 0 & 0 & 0 & 0 & 0 & 1 & 0 & 0  & 0  & 0  & 0  & 0  & 0  & 0  & 0  & 0  & 0  & 0  &  3.34   \\ \hline
\hline
\multicolumn{22}{|c|}{\textbf{Goa}}                                                                              \\ \hline
\textit{Person contact} & 1 & 2 & 3 & 4 & 5 & 6 & 7 & 8 & 9 & 10 & 11 & 12 & 13 & 14 & 15 & 16 & 17 & 18 & 19 & 20 & Avg. \\ \hline
\textit{Count}          & 0 & 0 & 0 & 0 & 0 & 0 & 0 & 0 & 0 & 0  & 0  & 0  & 0  & 0  & 0  & 0  & 0  & 0  & 0  & 0  &  0   \\ \hline
\hline
\multicolumn{22}{|c|}{\textbf{Gujarat}}                                                                              \\ \hline
\textit{Person contact} & 1 & 2 & 3 & 4 & 5 & 6 & 7 & 8 & 9 & 10 & 11 & 12 & 13 & 14 & 15 & 16 & 17 & 18 & 19 & 20 & Avg. \\ \hline
\textit{Count}          & 0 & 0 & 0 & 1 & 1 & 0 & 0 & 0 & 0 & 0  & 0  & 0  & 0  & 0  & 0  & 0  & 0  & 0  & 0  & 0  &  4.5   \\ \hline
\hline
\multicolumn{22}{|c|}{\textbf{Haryana}}                                                                  \\ \hline
\textit{Person contact} & 1 & 2 & 3 & 4 & 5 & 6 & 7 & 8 & 9 & 10 & 11 & 12 & 13 & 14 & 15 & 16 & 17 & 18 & 19 & 20 & Avg. \\ \hline
\textit{Count}          & 2 & 1 & 0 & 0 & 0 & 0 & 0 & 0 & 0 & 0  & 0  & 0  & 0  & 0  & 0  & 0  & 0  & 0  & 0  & 1  &  6   \\ \hline
\hline
\multicolumn{22}{|c|}{\textbf{Himanchal Pradesh}}                                                                              \\ \hline
\textit{Person contact} & 1 & 2 & 3 & 4 & 5 & 6 & 7 & 8 & 9 & 10 & 11 & 12 & 13 & 14 & 15 & 16 & 17 & 18 & 19 & 20 & Avg. \\ \hline
\textit{Count}          & 0 & 0 & 0 & 0 & 0 & 0 & 0 & 0 & 0 & 0  & 0  & 0  & 0  & 0  & 0  & 0  & 0  & 0  & 0  & 0  &  0    \\ \hline
\hline
\multicolumn{22}{|c|}{\textbf{Jammu and Kashmir}}                                                                              \\ \hline
\textit{Person contact} & 1 & 2 & 3 & 4 & 5 & 6 & 7 & 8 & 9 & 10 & 11 & 12 & 13 & 14 & 15 & 16 & 17 & 18 & 19 & 20 & Avg. \\ \hline
\textit{Count}          & 3 & 2 & 0 & 0 & 0 & 0 & 0 & 0 & 0 & 1  & 0  & 0  & 0  & 0  & 0  & 0  & 0  & 0  & 0  & 0  &  2.83  \\ \hline
\hline
\multicolumn{22}{|c|}{\textbf{Jharkhand}}                                                                              \\ \hline
\textit{Person contact} & 1 & 2 & 3 & 4 & 5 & 6 & 7 & 8 & 9 & 10 & 11 & 12 & 13 & 14 & 15 & 16 & 17 & 18 & 19 & 20 & Avg. \\ \hline
\textit{Count}          & 1 & 0 & 0 & 0 & 0 & 0 & 0 & 0 & 0 & 0  & 0  & 0  & 0  & 0  & 0  & 0  & 0  & 0  & 0  & 0  &  1   \\ \hline
\hline
\multicolumn{22}{|c|}{\textbf{Karnataka}}                                                                              \\ \hline
\textit{Person contact} & 1 & 2& 3 & 4 & 5 & 6 & 7 & 8 & 9 & 10 & 11 & 12 & 13 & 14 & 15 & 16 & 17 & 18 & 19 & 20 & Avg. \\ \hline
\textit{Count}          & 35 & 14 & 10 & 7 & 2 & 2 & 0 & 2 & 0 & 0  & 1  & 1  & 0  & 0  & 0  & 1  & 0  & 0  & 0  & 1  &  2.73   \\ \hline
\hline
\multicolumn{22}{|c|}{\textbf{Kerala}}                                                                              \\ \hline
\textit{Person contact} & 1 & 2& 3 & 4 & 5 & 6 & 7 & 8 & 9 & 10 & 11 & 12 & 13 & 14 & 15 & 16 & 17 & 18 & 19 & 20 & Avg. \\ \hline
\textit{Count}          & 7 & 5 & 1 & 3 & 0 & 4 & 0 & 0 & 0 & 0  & 0  & 0  & 0  & 0  & 0  & 0  & 0  & 0  & 0  & 0  &  2.8   \\ \hline
\hline
\multicolumn{22}{|c|}{\textbf{Ladakh}}                                                                              \\ \hline
\textit{Person contact} & 1 & 2 & 3 & 4 & 5 & 6 & 7 & 8 & 9 & 10 & 11 & 12 & 13 & 14 & 15 & 16 & 17 & 18 & 19 & 20 & Avg.  \\ \hline
\textit{Count}          & 2 & 0 & 0 & 0 & 0 & 0 & 0 & 0 & 0 & 0  & 0  & 0  & 0  & 0  & 0  & 0  & 0  & 0  & 0  & 0  &  0.5    \\ \hline
\hline
\multicolumn{22}{|c|}{\textbf{Madhya Pradesh}}                                                                              \\ \hline
\textit{Person contact} & 1 &2 & 3 & 4 & 5 & 6 & 7 & 8 & 9 & 10 & 11 & 12 & 13 & 14 & 15 & 16 & 17 & 18 & 19 & 20 & Avg. \\ \hline
\textit{Count}          & 2 & 1 & 1 & 0 & 0 & 0 & 0 & 0 & 0 & 0  & 1  & 0  & 0  & 0  & 0  & 0  & 0  & 0  & 0  & 0  &  3.6   \\ \hline
\hline
\multicolumn{22}{|c|}{\textbf{Maharashtra}}                                                                              \\ \hline
\textit{Person contact} & 1 & 2 & 3 & 4 & 5 & 6 & 7 & 8 & 9 & 10 & 11 & 12 & 13 & 14 & 15 & 16 & 17 & 18 & 19 & 20 & Avg. \\ \hline
\textit{Count}          & 2 & 2 & 2 & 1 & 2 & 0 & 0 & 0 & 0 & 0  & 1  & 0  & 0  & 0  & 0  & 0  & 0  & 0  & 0  & 0  &  3.7   \\ \hline
\hline
\multicolumn{22}{|c|}{\textbf{Manipur}}                                                                              \\ \hline
\textit{Person contact} & 1 & 2 & 3 & 4 & 5 & 6 & 7 & 8 & 9 & 10 & 11 & 12 & 13 & 14 & 15 & 16 & 17 & 18 & 19 & 20 & Avg. \\ \hline
\textit{Count}          & 0 & 0 & 0 & 0 & 0 & 0 & 0 & 0 & 0 & 0  & 0  & 0  & 0  & 0  & 0  & 0  & 0  & 0  & 0  & 0  &  0    \\ \hline
\hline
\multicolumn{22}{|c|}{\textbf{Mizoram}}                                                                              \\ \hline
\textit{Person contact} & 1 & 2 & 3 & 4 & 5 & 6 & 7 & 8 & 9 & 10 & 11 & 12 & 13 & 14 & 15 & 16 & 17 & 18 & 19 & 20 & Avg. \\ \hline
\textit{Count}          & 0 & 0 & 0 & 0 & 0 & 0 & 0 & 0 & 0 & 0  & 0  & 0  & 0  & 0  & 0  & 0  & 0  & 0  & 0  & 0  &  0    \\ \hline
\hline
\multicolumn{22}{|c|}{\textbf{Odisha}}                                                                              \\ \hline
\textit{Person contact} & 1 & 2& 3 & 4 & 5 & 6 & 7 & 8 & 9 & 10 & 11 & 12 & 13 & 14 & 15 & 16 & 17 & 18 & 19 & 20 & Avg. \\ \hline
\textit{Count}          & 0 & 1 & 0 & 0 & 0 & 0 & 0 & 0 & 0 & 0  & 0  & 0  & 0  & 0  & 0  & 0  & 0  & 0  & 0  & 0  &  1   \\ \hline
\hline
\multicolumn{22}{|c|}{\textbf{Puducherry}}                                                                              \\ \hline
\textit{Person contact} & 1 & 2 & 3 & 4 & 5 & 6 & 7 & 8 & 9 & 10 & 11 & 12 & 13 & 14 & 15 & 16 & 17 & 18 & 19 & 20 & Avg. \\ \hline
\textit{Count}          & 0 & 0 & 0 & 0 & 0 & 0 & 0 & 0 & 0 & 0  & 0  & 0  & 0  & 0  & 0  & 0  & 0  & 0  & 0  & 0  &  0   \\ \hline
\hline
\multicolumn{22}{|c|}{\textbf{Punjab}}                                                                              \\ \hline
\textit{Person contact} & 1 & 2 & 3 & 4 & 5 & 6 & 7 & 8 & 9 & 10 & 11 & 12 & 13 & 14 & 15 & 16 & 17 & 18 & 19 & 20 & Avg. \\ \hline
\textit{Count}          & 7 & 0 & 1 & 0 & 0 & 1 & 1 & 0 & 0 & 0  & 0  & 0  & 0  & 0  & 0  & 0  & 0  & 0  & 0  & 1  & 3.91   \\ \hline
\hline
\multicolumn{22}{|c|}{\textbf{Rajasthan}}                                                                              \\ \hline
\textit{Person contact} & 1 & 2 & 3 & 4 & 5 & 6 & 7 & 8 & 9 & 10 & 11 & 12 & 13 & 14 & 15 & 16 & 17 & 18 & 19 & 20 & Avg. \\ \hline
\textit{Count} & 4 & 3 & 0 & 1 & 0 & 0 & 0 & 0 & 0 & 0  & 0  & 0  & 0  & 0  & 0  & 0  & 0  & 0  & 0  & 0  &  1.75    \\ \hline
\hline
\multicolumn{22}{|c|}{\textbf{Tamil Nadu}}                                                                              \\ \hline
\textit{Person contact} & 1 & 2 & 3 & 4 & 5 & 6 & 7 & 8 & 9 & 10 & 11 & 12 & 13 & 14 & 15 & 16 & 17 & 18 & 19 & 20 & Avg.  \\ \hline
\textit{Count}          & 6 & 0 & 2 & 0 & 0 & 0 & 1 & 0 & 0 & 0  & 0  & 0  & 0  & 1  & 0  & 0  & 0  & 0  & 0  & 0  &  3.30   \\ \hline
\hline
\multicolumn{22}{|c|}{\textbf{Telangana}}                                                                              \\ \hline
\textit{Person contact} & 1 & 2 & 3 & 4 & 5 & 6 & 7 & 8 & 9 & 10 & 11 & 12 & 13 & 14 & 15 & 16 & 17 & 18 & 19 & 20 & Avg. \\ \hline
\textit{Count}          & 5 & 1 & 0 & 1 & 0 & 0 & 0 & 0 & 0 & 0  & 0  & 0  & 0  & 0  & 0  & 0  & 0  & 0  & 0  & 0  &  1.16    \\ \hline
\hline
\multicolumn{22}{|c|}{\textbf{Uttar Pradesh}}                                                                              \\ \hline
\textit{Person contact} & 1 & 2 & 3 & 4 & 5 & 6 & 7 & 8 & 9 & 10 & 11 & 12 & 13 & 14 & 15 & 16 & 17 & 18 & 19 & 20 & Avg.  \\ \hline
\textit{Count}          & 7 & 0 & 1 & 1 & 2 & 0 & 1 & 0 & 0 & 0  & 0  & 0  & 0  & 0  & 0  & 0  & 0  & 0  & 0  & 0  &  2.58    \\ \hline
\hline
\multicolumn{22}{|c|}{\textbf{Uttrakhand}}                                                                              \\ \hline
\textit{Person contact} & 1 & 2 & 3 & 4 & 5 & 6 & 7 & 8 & 9 & 10 & 11 & 12 & 13 & 14 & 15 & 16 & 17 & 18 & 19 & 20 & Avg. \\ \hline
\textit{Count}          & 0 & 0 & 0 & 0 & 0 & 0 & 0 & 0 & 0 & 0  & 0  & 0  & 0  & 0  & 0  & 0  & 0  & 0  & 0  & 0  &  0    \\ \hline
\hline
\multicolumn{22}{|c|}{\textbf{West Bengal}}                                                                              \\ \hline
\textit{Person contact} & 1 & 2 & 3 & 4 & 5 & 6 & 7 & 8 & 9 & 10 & 11 & 12 & 13 & 14 & 15 & 16 & 17 & 18 & 19 & 20 & Avg. \\ \hline
\textit{Count}          & 0 & 1 & 1 & 0 & 0 & 0 & 0 & 0 & 0 & 0  & 0  & 0  & 0  & 0  & 0  & 0  & 0  & 0  & 0  & 0  &  2.5   \\ \hline
\hline
\multicolumn{19}{|c|}{\textbf{India}}& \multicolumn{3}{|c|}{\textbf{Avg.} = 1.79}                                                                              \\ \hline
\end{tabular}
}
\label{tab2}
\end{table*}

\vspace{-0.5cm}
\subsection{Epidemic second parameter: Case fatality rate}\label{fatality}
\begin{figure}[h]
\centering
\includegraphics[scale=1]{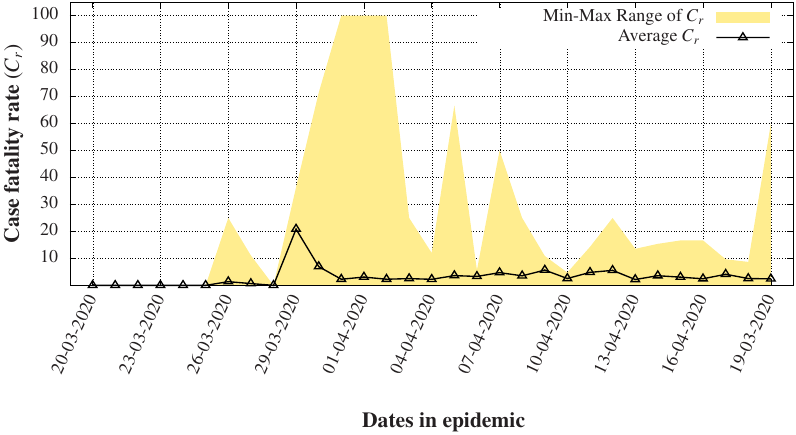}
\caption{Illustration of $C_r$ with time in India.}
\label{crgraph}
\end{figure}

\textcolor{black}{The case fatality rate, denoted by  $C_r$, provides an estimation of the number of the person who dies among the total infected person from disease during its spread in a particular area (state/country). It defines as,}
\begin{definition} Case fatality rate of a given region is the ratio of \textcolor{black}{the total number of persons who died from the epidemic to the total number of the persons who are infected, \textit{i.e.},}
\begin{equation} 
 C_r = \frac{\text{Total death from the epidemic}}{\text{Total number of persons who are infected}}\times 100.
\end{equation}
\end{definition}
\noindent $\bullet$ \textbf{Importance of $C_r$}: The reproduction number can only give the idea of the rate of disease spread, but cannot estimate how many persons can lose their life in the entire cycle of the spread. Therefore, the case fatality rate is crucial for such estimation. An epidemic like measles with $R_o$ around $18$ has a lower case fatality rate to that of Ebola with $R_o$ around $1.5$~\cite{blog}. Thus, the case fatality rate plays a vital role in estimating the actual effect of a disease on humanity. For example, Ebola with lesser $R_o$ is more severe than the measles. \\
$\bullet$ \textbf{$C_r$ in given dataset}: Fig.~\ref{crgraph} illustrates the value of $C_r$ for the given day. It also shows the minimum and maximum $C_r$ \textcolor{black}{on the states of India each day. We conclude from the result that the difference between the minimum and maximum value of $C_r$ is very high, which indicates that an epidemic is very high in some state while others are free. The result also illustrates the minimum value of $C_r$ increases with time. 
It indicates the epidemic is spreading in India at a slower rate.}  Table~\ref{cfr} illustrates the case fatality rate of different states of India from COVID-19. The fatality rate of Himanchal Pradesh is highest, \textit{i.e.,} $1.79$. High fatality rate indicates that a large number of infected persons die in these states. However, the fatality rate also increases if the number of infected patients is low. It seems right in the case of Himanchal Pradesh too, as the reported cases in the state are only $39$ out of which one is deceased. The fatality rate of few states is zero as there is no casualty reported so far in them like Andaman and Nicobar,  Assam, Chandigarh, \textit{etc}. The overall fatality rate of India reported up till April, $18$ $2020$ is $0.34$, which \textcolor{black}{indicates person infection rate} is low in India and they are gaining significant recovery from the COVID-19.

 \begin{table}[h]
\centering
\scriptsize
\rowcolors{2}{gray!35}{white}
\caption{Illustration of case fatality rate in states of India.}
\begin{tabular}{|c|c|c|c|}
\hline
\textbf{State/UT}   & \textbf{\begin{tabular}[c]{@{}c@{}}Fatality \\ rate (in\%)\end{tabular}} & \textbf{State/UT}  & \textbf{\begin{tabular}[c]{@{}c@{}}Fatality\\ rate (in\%)\end{tabular}} \\ \hline \hline
Andaman and Nicobar & 0                                                                 & Andhra Pradesh     & 0                                                                \\ \hline
Arunachal Pradesh   & 0                                                                 & Assam              & 0                                                                \\ \hline
Bihar               & 0.71                                                              & Chandigarh         & 0                                                                \\ \hline
Chattisgarh         & 0                                                                 & Dadra Nagar Haveli & 0                                                                \\ \hline
Delhi               & 0.044                                                             & Goa                & 0                                                                \\ \hline
Gujrat              & 0.17                                                              & Haryana            & 0                                                                \\ \hline
Himanchal Pradesh   & 2.56                                                              & Jammu and Kashmir  & 0.246                                                            \\ \hline
Jharkhand           & 0                                                                 & Karnataka          & 2.48                                                                \\ \hline
Kerla               & 0.46                                                              & Ladakh             & 0                                                                \\ \hline
Madhya Pradesh      & 0.126                                                            & Maharashtra        & 0.195                                                            \\ \hline
Manipur             & 0                                                                 & Mizoram            & 0                                                                \\ \hline
Odisha              & 1.21                                                              & Puducherry         & 0                                                                \\ \hline
Punjab              & 0.72                                                              & Rajasthan          & 0.052                                                             \\ \hline
Tamil Nadu          & 0.061                                                             & Telangana          & 0.11                                                                \\ \hline
Tripura             & 0                                                                 & Uttar Pradesh      & 0.069                                                            \\ \hline
Uttrakhand       & 0                                                                    & West Bengal         & 0.874                                                                \\ \hline
\multicolumn{3}{|c|}{\textbf{India}}                                                                         & \textbf{0.34 (in\%)}                                                   \\ \hline
\end{tabular}
\label{cfr}
\end{table}

\subsection{ Epidemic third parameter: State of given region}\label{state1}
\begin{figure}[h]
\centering
\includegraphics[scale=1.0]{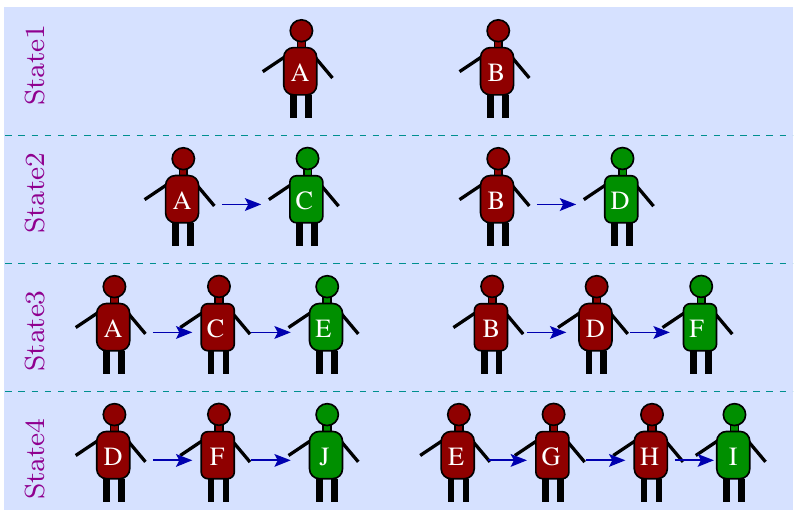}
\caption{An example scenario representing four transmission state of an epidemic.}
\label{example1}
\end{figure}

Measures such as the social and physical distancing \textcolor{black}{reaccelerate the} spread of the disease, infections by stopping the chains of transmission from one person to another. These preventive measures reduce the contamination from the persons or surfaces while encouraging connections with family members. The spread of the infection is broadly categorised in the following four infection transmission states~\cite{gurley2007person}:

\begin{enumerate}
 \item \textbf{State 1 (No contact transmission):} In this transmission state, the infection from an infected person does not proceed further to any other person. The infected person remains confined with the disease-causing virus without further dissipation. For an epidemic disease like COVID-19, this state is the most crucial state to control the spread of coronavirus through quarantine and social isolation. However, several infected persons in this state are unaware of their infection. Thus, passing the infection to the person in the contact and accelerate the growth of the disease. For example, the topmost layer in Fig.~\ref{example1} demonstrates two persons (\textit{i.e.,} A and B) are infected with coronavirus and there is no any further transmission. This may result due to the complete isolation of both A and B.
 
 \item \textbf{State 2 (Local transmission):} The isolation compromise and delay in the detection of infection in a person leads this transmission state. The infected person transfers the infection to the person in its close contact. This type of transmission is usually observed in the case when the infected person comes in close contact with a person like a friend or a family member. At this stage, it is easy to trace spread and quarantine people.  The quarantine of the infected person must be strictly followed to avoid this state to enter into the next state. The second layer in Fig.~\ref{example1} indicates the transmission is from infected person A to C and B to D.  
 
 \item \textbf{State 3 (Untraceable transmission):} The unaltered social contact between the persons results in transmission state third. In this state, the concept of transmitting and being transmitted is followed. Here, an infected person transfers the infection to another person who in turn transfers to the third person. The third stage is also when the source of the infection is untraceable. The infected people in this stage are identified who haven’t had travel history. This state, in case of COVID-19, is controlled by adopting a hard mechanism like lockdowns over the entire country. The third stage in Fig.~\ref{example1} illustrates State 3. In this state, A transfers infection to C that further propagates to E. Similar transmission is observed among B, D, and F. 
 
 \item \textbf{State 4 (Source missing transmission):} This is the final infection transmission state where the source of infection is unidentified. It can be understood by an example where a person is infected in spite of not coming in close contact with an infected person and had no travel history from an infected country. This transmission state is most severe and leads to infection spread to all the persons in a particular area where the infection persists. The last layer in Fig.~\ref{example1} illustrates the fourth state of transmission, where, the source is unidentified.
\end{enumerate}

\begin{figure}[h]
\centering
\includegraphics[scale=1]{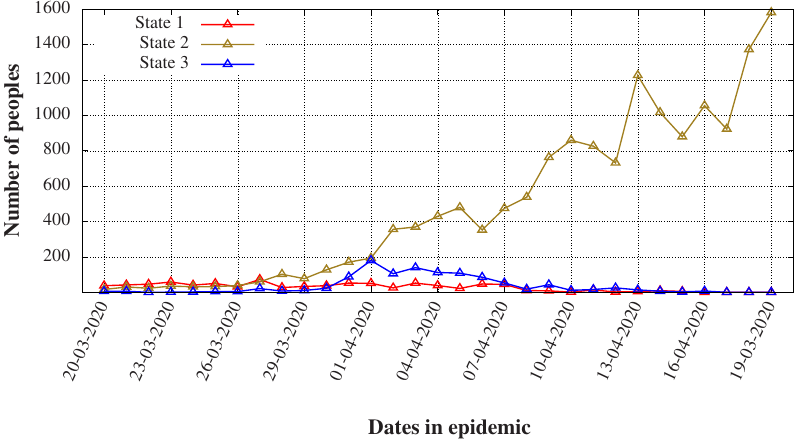}
\caption{Illustration of state 1, state 2, and state 3 persons.}
\label{statefigure}
\end{figure}

\noindent $\bullet$ \textbf{Algorithm for identify state of epidemic}: Algorithm~$1$ illustrates the steps involved in the identification of the infected population in the different transmission state. In the algorithm, we introduce a variable ${infection\_tramission}$ that predicts the chance of a person to lie in a particular transmission state depending upon its values (\textit{i.e.,} $0$,  $1$, and more).  ${infection}_{transmission}=0$ indicates that there is no person-to-person infection transmission and  ${infection}_{transmission} = 1$ indicate that infection transmission is in State $2$. Further, if the infection is transmitted from the first person to second person and then to the third or the source is untraceable indicates the transmission State $3$. If the transmission state does not belong to the above three, then the infection transmission State $4$ is reached. At State $4$, it is assumed that the entire population is equally likely to be infected. The estimation of infection transmission State $4$ is out of scope from this algorithm. To hamper the infection transmission, the measures for the peoples includes the closure of non-essential services, work from home, avoid social gathering, isolating vulnerable groups, \textit{etc}. These measures are successful if peoples follow conjunctive measures such as hand washing frequently.
\SetAlFnt{\small}
\begin{algorithm}[h]
\caption{\textbf{Epidemic state identification}}
\KwIn{ Population $M$ of a particular area (state/country)\;} 
\KwOut{States in which all the infected population lie\;}
\nonl /*Variable initialization*/\\
${Set}_1$=\{\}, ${Set}_2$=\{\}, and ${Set}_3$=\{\}.\\
${infection\_tramission}\leftarrow0$.\\
Identify the infectious population $I$.\\
\For{ $i\leftarrow$ 1 to $\|I\|$}{
\If{${infection\_tramission}_i == 0$}{
${Set}_1\leftarrow {Set}_1.\text{append}(\text{person}_i)$.
}
\ElseIf{${infection\_tramission}_i == 1$}{
${Set}_2\leftarrow {Set}_2.\text{append}(\text{person}_i)$.
}
\Else{
${Set}_3\leftarrow {Set}_3.\text{append}(\text{person}_i)$.
}}
The resultant sets ${Set}_1$, ${Set}_2$, and ${Set}_3$ contain the infected person in different transmission state, \textit{i.e.}, State $1$, State $2$, and State $3$, respectively.\\
\label{algorithm1}
\end{algorithm}\\
\noindent $\bullet$ \textbf{State in given dataset:} Fig~\ref{statefigure} illustrates the number of persons each state with time. As we expect, the number of cases are registered more with time. An interesting observation from this result is that the number of persons in state 2 increases sharply then state 1. This is because of the state 1 person infected to state 2 with a high value of $R_o$. Table~\ref{state} illustrates the different state of India, having different person indulge in all four transmission states. Due to a large number of cases in the state of Maharashtra, the number of persons in transmission state four is high. The table is created using the data available at~\cite{data} (April $18$ $2020$). An important conclusion from the table is that India is in Stage 2 ($90.11\%$) and shifting in Stage 3 ($5.13\%$).
\begin{table}[h]
\centering
\scriptsize
\rowcolors{2}{gray!35}{white}
\caption{Identify the state of transmission.}
\begin{tabular}{|c|c|c|c|}
\hline 
\textbf{State/UT}   & \textbf{State1} & \textbf{State2} & \textbf{State3} \\ \hline
\hline
Andaman and Nicobar & 5               & 1               & 2              \\ \hline
Andhra Pradesh      & 18              & 731             & 64             \\ \hline
Arunachal Pradesh   & -               & -               & 1              \\ \hline
Assam               & 1               & 10              & 25             \\ \hline
Bihar               & 24              & 111             & 8              \\ \hline
Chandigarh          & 12              & 14              & 1              \\ \hline
Chattisgarh         & 6               & 29              & 1              \\ \hline
Dadra Nagar Haveli  & -               & -               & -               \\ \hline
Delhi               & 22              & 2192            & 34              \\ \hline
Goa                 & 7               & -               & -               \\ \hline
Gujrat              & 121             & 2280            & 6               \\ \hline
Haryana             & 29              & 219             & 16              \\ \hline
Himanchal Pradesh   & 3               & 21              & 15               \\ \hline
Jammu and Kashmir   & 21              & 380             & 6                \\ \hline
Jharkhand           & 2               & 47              & -              \\ \hline
Karnataka           & 89              & 224              & 114                \\ \hline
Kerla               & 259             & 156              & 22              \\ \hline
Ladakh              & 7               & -               & -               \\ \hline
Madhya Pradesh      & 17              & 1568            & 2               \\ \hline
Maharasthra         & 82              & 5547            & 20              \\ \hline
Manipur             & 1               & -               & 1                \\ \hline
Odisha              & 19               & 64               & 1        \\ \hline
Puducherry          & 1               & 2              & 4                           \\ \hline
Punjab              & 30              & 223               & 25                     \\ \hline
Rajasthan           & 96              & 1811               & 28                     \\ \hline
Tamil Nadu          & 70              & 1046               & 513                      \\ \hline
Telangana           & 29               & 888               & 26                     \\ \hline
Tripura             & -               & 2               & -                         \\ \hline
Uttar Pradesh       & 25              & 1264              & 160                        \\ \hline
Uttrakhand          & 5               & 37               & 4                         \\ \hline
West Bengal         & 16               & 438               & 2                 \\ \hline 
INDIA (Total)        & 1017               & 19305               & 1101                 \\  
INDIA (in \%)        & \textbf{4.74}               & \textbf{90.11}               & \textbf{5.13}                 \\ \hline 
\end{tabular}
\label{state}
\end{table}

\section{Epidemic Models and Contact tracking}\label{model1}
Data models help in quantifying the damage caused by a disease like epidemic COVID-19. The model simulates the spread of infection from one person to another and its geographical spread on the entire population. The models also help in identifying the time after which the epidemic end or its effect can be reduced. Data models predict the number of population that is infectious, susceptible and recovered in the near future. In this section, we discuss the exiting epidemic models and how to use on the dataset. 
\subsection{Susceptible, Infectious, and Recovered (SIR) model}\label{sir}
\begin{figure}[h]
\centering
\includegraphics[scale=1.00]{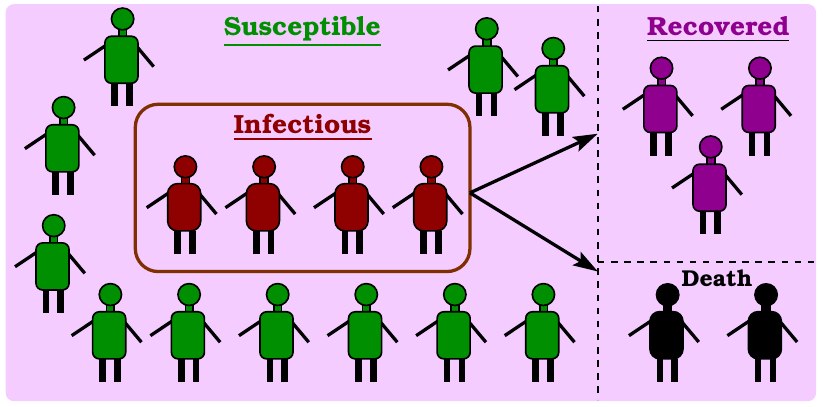}
\caption{Illustration of susceptible, infectious, and recovered persons of a particular area.}
\label{otherfig1}
\end{figure}

The most generalised framework for such simulation is based on Susceptible, Infectious, and Recovered (SIR) model~\cite{7029011}. The SIR model is a three-phase model where a person partially stays in a phase and switches to another with a specific course of action.  Let an epidemic spread of disease covers the entire population of $M$ persons having Susceptible $S$ and Infectious $I$ with Recovered $R$ persons, then $M=S+I+R$. The phases of SIR models are illustrated in Fig.~\ref{otherfig1}. The description of each phase is as follows.

\noindent \textbf{Susceptible:} All the person in a particular area who are not infected with the disease and have no reported immunity against the infection are categorised as Susceptible (S). The absence of immunity in a person is the result of no prior exposure to the infection, non-vaccinated against such infection, and no confirm disease resistance. The susceptible person moves to the infectious phase by social contact with an infected person.

\noindent \textbf{Infectious:} During the disease outbreak like COVID-19, some person come in close contact with the infected person and got infected afterwards, forming a group of persons termed as Infectious (I). These persons are solely responsible for the spread of the disease in the whole community. The quarantine of such infected persons preserves the community from being infectious. Social isolation from such a person is the only remedy against any epidemic outbreak like COVID-19.  

\noindent \textbf{Recovered:} Every disease like epidemic COVID-19 has a specific cycle for infection persistence. After this cycle, the infected person recovers from the infection with proper treatment. However, due to the severe spread of infection in the human body and inadequate cure, the infection can result in death. The summation of total recovered and total dead is termed as Recovered (R) or sometimes removed. The persons that are in this category are free from further infection.  

There is always a phase of transmission among the persons, \textit{i.e.,} from susceptible to infectious and infectious to susceptible. The transmission of the disease changes the number of persons in a phase over time. The SIR model can be represented using~\cite{8957067} as follows
\begin{align}\label{R}
 R=&\alpha_2 I-\gamma I,\\\label{I}
 I=&\alpha_1 S,\\\label{eq3}
 M=&S\cup I\cup R,
\end{align}
where, $\alpha_1$ and $\alpha_2$ are the force of infection and recovery rate, respectively. $\gamma$ denotes the death rate and $\cup$ is the union operator.
Eq.~\ref{eq3} holds when the entire population of a specified area is susceptible to the epidemic outbreak. 

Force of infection $\alpha_1$ is the rate at which the susceptible person becomes infectious on the close contact with the infected person. Recovery rate $\alpha_2$ is the rate at which an infected person recovers from the infection using the proper treatment. The force of infection is not a constant quantity and is proportional to the transmission rate $\tau$. The transmission rate $\tau$ is the product of the rate of contact $r_c$ and the probability of transmission $p_t$ from susceptible to infectious phase. $\tau$ is defined as
\begin{equation}\label{transmission}
 \tau = r_c \times p_t.
\end{equation}
Using Eq.~\ref{transmission} we can define force of infection $\alpha_1$ as
\begin{equation}\label{force}
 \alpha_1(I)=\tau \Big\{\frac{I}{M} \Big\}\times 100,
\end{equation}
where, $\{\frac{I}{M}\}\times 100$, represents percentage of population, which is infectious among the total population in a particular area.

To capture the population change in a particular phase, SIR model adopts differential equations. A differential equation captures the progression of the disease.
The rate of change in the susceptible population (S), infectious population (I), and recovered population (R) as defined as follows
\begin{align}
 \frac{\mathrm{d}S}{\mathrm{d}t} &= -\alpha_1(I)S.\\
 \frac{\mathrm{d}I}{\mathrm{d}t} &=\alpha_1(I)S-\alpha_2I.\\\label{recovery}
 \frac{\mathrm{d}R}{\mathrm{d}t} &= \alpha_2I.
\end{align}

An essential step after estimating the rate of change in $S$, $I$, and $R$ using the differential equation, is to determine the equilibrium state of the SIR model. To describe the epidemic methodology, two equilibria exist side-by-side, \textit{i.e.,} Disease Free Equilibrium  and Endemic Equilibrium~\cite{pardoux2020deviations}. The Disease Free Equilibrium comes into the picture when there is no person left with the infection in a particular area. However, in the case of Endemic Equilibrium there always persists an infectious individual. It requires a constant supply of susceptible person to maintain the equilibrium. Both the equilibrium persists when the value of $R_o$ is less than $1$. The SIR model reaches an equilibrium state if the rate of transmission form susceptible to infectious is equal to infectious to recover. Algorithm~$2$ illustrates the process of estimating the time $\mathbf{t}$ for which an epidemic persists using Susceptible, Infectious, and Recovered model.

\SetAlFnt{\small}
\begin{algorithm}[h]
\caption{\textbf{Susceptible, Infectious, and Recovered}}
\KwIn{Population $M$ of a particular area facing epidemic\;}
\KwOut{Time $T$ for which the epidemic persist\;}
\nonl /*Variable initialization*/\\
$\alpha_1 \leftarrow 0, \alpha_2\leftarrow0, \tau\leftarrow0,\text{ and } r_c\leftarrow0$.\\
Time $\mathbf{t}\leftarrow 0$ and change in time $\Delta t$.\\
Calculate reproduction number $R_o$. \\

\While{$R_o<1$ or  $\frac{\mathrm{d}S}{\mathrm{d}t} = 
 \frac{\mathrm{d}I}{\mathrm{d}t} =
 \frac{\mathrm{d}R}{\mathrm{d}t} = 0$}{

Estimating transmission rate $\tau$\\
\hspace{5pt} Calculate rate of contact ($r_c$) among the persons.\\
\hspace{5pt} Calculate probability of transmission ($p_t$). \\
\hspace{7pt} $\tau\leftarrow r_c \times p_t$.\\
Estimate \textit{force of infection} $\alpha_1$ using Eq.~\ref{force}.\\
Calculate $R$ and $I$ for given $S$.\\
\hspace{7pt}$I=\alpha_1 S$ and $R=\alpha_2 I-\gamma I$ using Eqs.~\ref{I} and~\ref{R}. \\
Calculate the derivatives $\frac{\mathrm{d}S}{\mathrm{d}t}$, $\frac{\mathrm{d}I}{\mathrm{d}t}$, and $\frac{\mathrm{d}R}{\mathrm{d}t}$. \\
\hspace{5pt} $\mathbf{t} \leftarrow \mathbf{t} + \Delta t$.
}
Either Disease Free or Endemic Equilibrium is reached.\\
The epidemic will be ended after $\mathbf{t}$ time.
\label{algorithm2}
\end{algorithm}

\begin{figure}[h]
\centering
\includegraphics[scale=0.43]{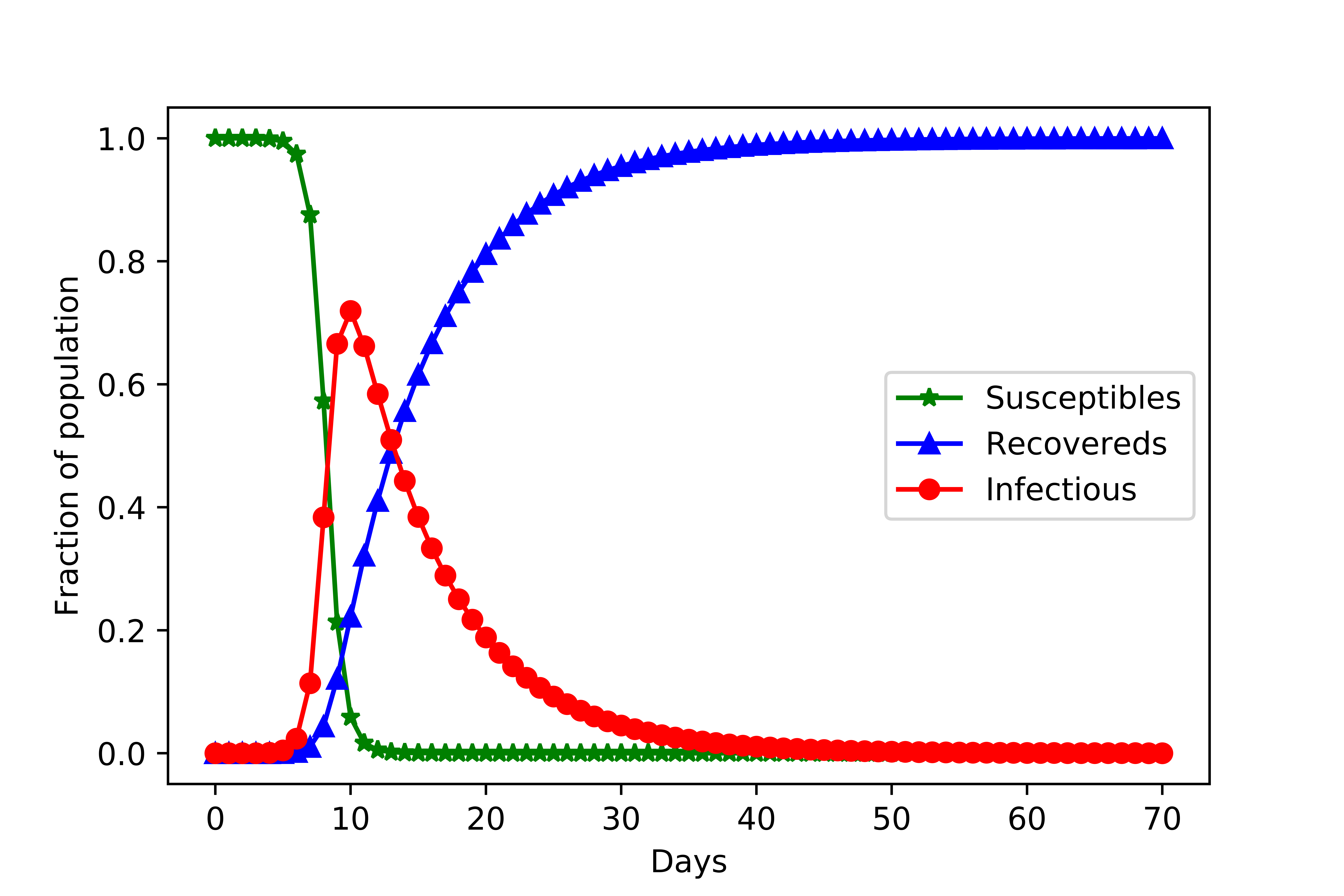}
\caption{Illustration of SIR model on coronavirus dataset of India up till April $18$ $2020$.}
\label{sir_India}
\end{figure}

\noindent $\bullet$ \textbf{SIR model on the given dataset:}
Fig~\ref{sir_India} illustrates the relationship between susceptible, infectious, and recovered cases in India up till April $18$ $2020$. The figure demonstrates that infectious persons grow up to a limiting value, where recovery starts at a higher pace. It compels most of the people to come out of the susceptible, and the outbreak is about to die. During the construction of the result, we have taken the value of reproduction number $R_o=1.79$, \textit{i.e.,} current status for India with a total  $12,079$ active cases. We can observe that if the rate of new case generation in India is decreasing or not exceeding the current rate, then the COVID-19 breakout will be ended in the next $70$ days.

\subsection{Susceptible Infectious (SI) model}
Chermack and McKinsey~\cite{ranney2013correlates} proposed the Susceptible Infectious (SI) model. The authors assumed that the total population in a particular area during epidemic breakout belongs either to susceptible or infectious. Apart from the SIR model, they only consider the susceptible and infectious population without considering the recovered cases in a particular location. Let at time $t=0$ (time at which the estimation begins), the number of infectious and susceptible peoples are represented as $I_o$ and $S_o$, respectively. Let $\tau$ represents the transmission rate given in Eq.~\ref{transmission}. Then, the rate of change in susceptible and infectious is given as
 \begin{align}\label{rates}
 \frac{\mathrm{d}S}{\mathrm{d}t}= -\tau S I.\\\label{ratei}
 \frac{\mathrm{d}I}{\mathrm{d}t}= \tau I(1-I). 
 \end{align}
Upon further calculation as given in~\cite{ranney2013correlates}, the number of infectious during an epidemic breakdown can be obtained as follows
\begin{equation}
 I=\frac{I_o e^{\tau t}}{1-I_o + I_o e^{\tau t}}.
\end{equation}

\begin{figure}[h]
\centering
\includegraphics[scale=0.43]{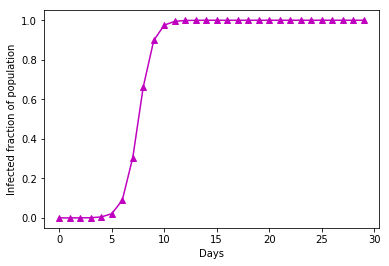}
\caption{Illustration of infectious population using SI model on coronavirus dataset of India up till April $18$ $2020$.}
\label{si_India}
\end{figure}

\noindent $\bullet$ \textbf{SI model on the given dataset:}
Fig.~\ref{si_India} illustrates the fraction of the infectious population for past $30$ days of the epidemic (COVID-19) breakout in India. The result shows that the curved we obtained is an S-shaped curve for infectious. The curve grows exponentially from the early stage of infection in the country the infection starts spreading. During the result estimation, we have taken the value of reproduction number $R_o=1.79$, \textit{i.e.,} current status for India with a total  $12,079$ active cases as on April $18$ $2020$. We can observe that if the rate of new case generation in India is increasing, which in turn increases the infectious population in the country. This result seems to be an alarming situation for taking more preventive measures to hamper the growing cases in the country.

\subsection{Susceptible Infectious and Susceptible (SIS) model }\label{sis}
Susceptible Infectious and Susceptible (SIS) model is an extension of the SI model~\cite{shi2008sis}. The authors assumed that the people who recovered from the infection are still susceptible. It means the those who recovered from infection during an epidemic has not developed immunity against it. The rate of change in susceptible is given as
\begin{equation}
 \frac{\mathrm{d}S}{\mathrm{d}t}= -\tau S I + \alpha I,
\end{equation}
where, $\tau$ and $\alpha$ represents the transmission rate and fraction of the infected population, respectively. The term $\tau S I$ can be understood as an average infection transmitted by an individual through sufficient contacts for transmitting the infection to $\tau M$ susceptible population, for $M$ as the population size of a particular area. The probability that a given person can infect other is $S/M$ in this case. Therefore, the infection transmission per unit time is $\tau S$ by a single individual. The rate of change in Infectious population is as follows

\begin{equation}
  \frac{\mathrm{d}I}{\mathrm{d}t}= \tau S I - \alpha I.
\end{equation}

Using the assumption that the epidemic could infect the entire population such that $M=S+I$. Therefore, we can obtain
\begin{align}\nonumber
 \frac{\mathrm{d}I}{\mathrm{d}t}=& \tau I (M-I)  - \alpha I,\\
 =& (\tau M -\alpha)I -\tau I^{2}.
\end{align}
Solving for $\frac{\mathrm{d}I}{\mathrm{d}t}=0$, we can obtain two possible equilibria, \textit{i.e,} at $I=0$ and $I=M-\frac{\alpha}{\tau}$. When we achieve the equilibrium then the epidemic is about to end.

\begin{figure}[h]
\centering
\includegraphics[scale=0.43]{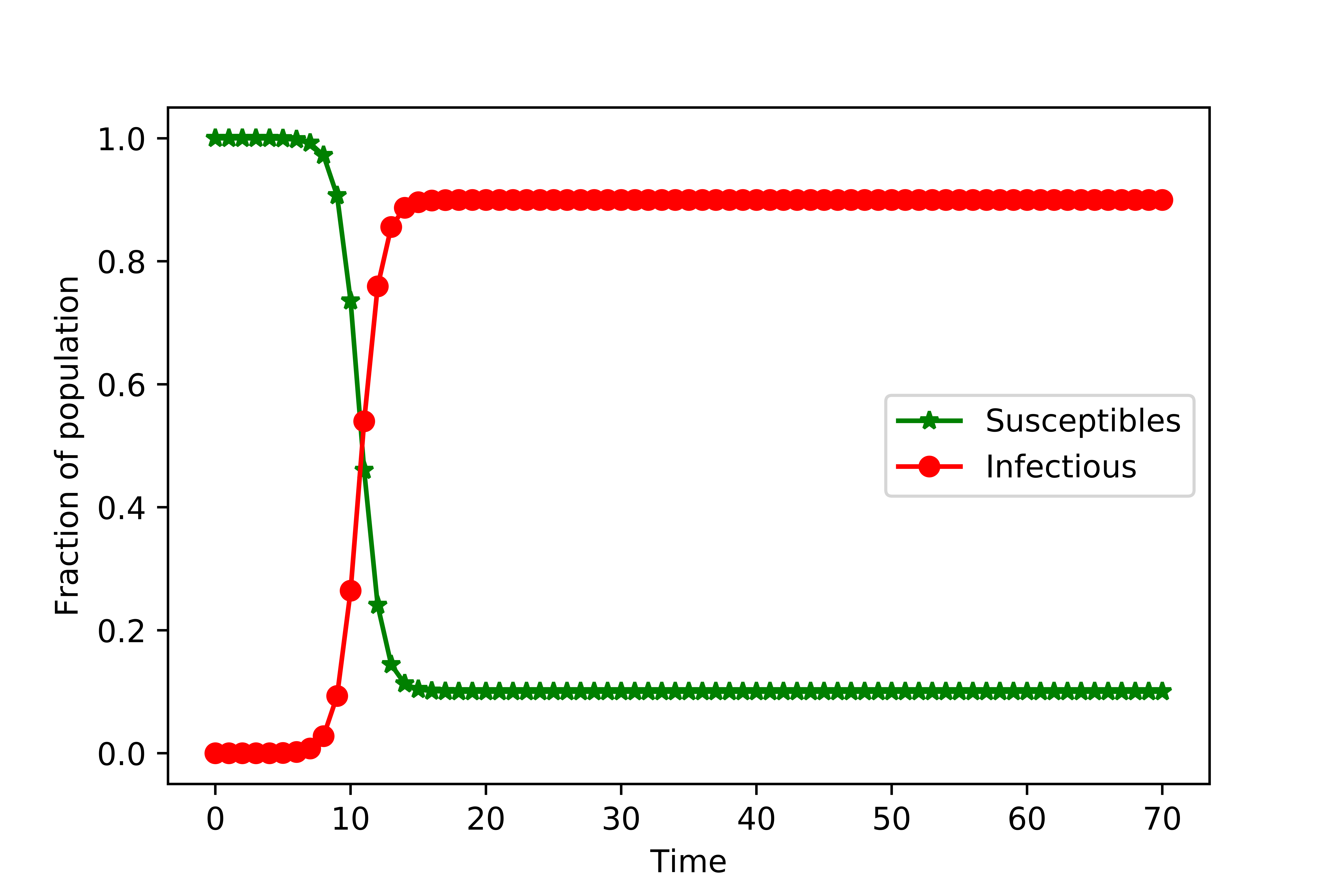}
\caption{Illustration of infectious and susceptible population curve of SIS model on dataset of India upto April $18$ $2020$.}
\label{sis_India}
\end{figure}

\noindent $\bullet$ \textbf{SIS model on the given dataset:}
The result illustrated in Fig~\ref{sis_India} depicts the relationship between susceptible and the infectious fraction of cases in India up till April $18$ $2020$. The result is estimated by taking the value of the reproduction number as  $R_o=1.79$, \textit{i.e.,} current status for India with a total   $12,079$ active cases up to April $18$ $2020$. Similarly, to previous observations in the SIR model in Fig.~\ref{sir_India}, here also, we can observe that if the rate of new case generation in India is decreasing or not exceeding the current rate, then the COVID-19 breakout will be ended in the next $70$ days.

\section{Results using empirical analysis}\label{res1}
This work carried experiment using empirical analysis of population size $1000$ persons. We consider the value of the reproduction number ($R_o$) as $1.79$. This value is taken as the average reproduction number for COVID-19 in India is $1.79$. 
The empirical analysis in this paper is based on the following formula derived from SIR model~\cite{ma2003stability}.
\begin{equation}
 I_{total}=I_o + r_c \Big( \frac{I}{M} \Big)\times 100 \times S \times p_t,
\end{equation}
where, $I_o$ is the initial infected population, $r_c$ is the rate of contact, and $p_t$ is the transmission probability.

\subsubsection{Impact of Infectious and Susceptible}
\begin{figure}[h]
\centering
\includegraphics[scale=1]{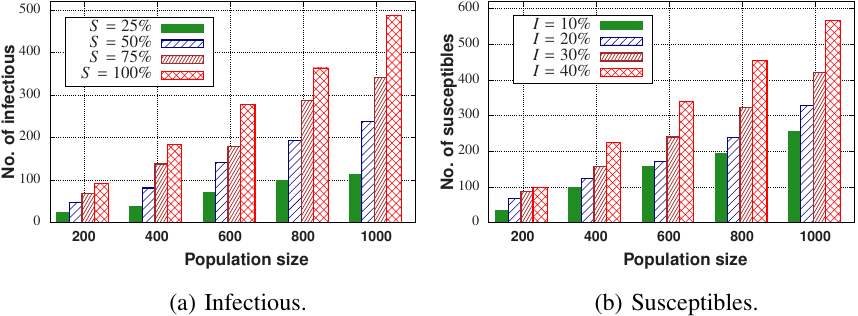}  
\caption{Impact of number of population on infectious and susceptible, respectively.} 
\label{is1}
\end{figure}

This work estimates the infectious population when the susceptible increase from $25\%$ to $100\%$, with the rate of contact $r_c=1$ and the probability of transmission $p_t=0.3$. Part (a) of Fig.~\ref{is1} illustrates that the number of infected persons increases with the increase in the \% susceptible. If the susceptible reaches to $100\%$ on population size $1000$, then the infection can spread up to $487$ persons. Similarly, part (b) of Fig.~\ref{is1} illustrates that the increment in the infectious accelerates the increment in the susceptible population and reaches $564$ when population size is $1000$ and infectious is $40\%$. These results illustrate that infectious and susceptible are directly related to each other and increases in a similar manner. Therefore, the effect of an epidemic such as COVID-19 can be reduced by decreasing either the susceptible or infectious population. 

\begin{figure*}[ht]
\centering
\includegraphics[scale=1]{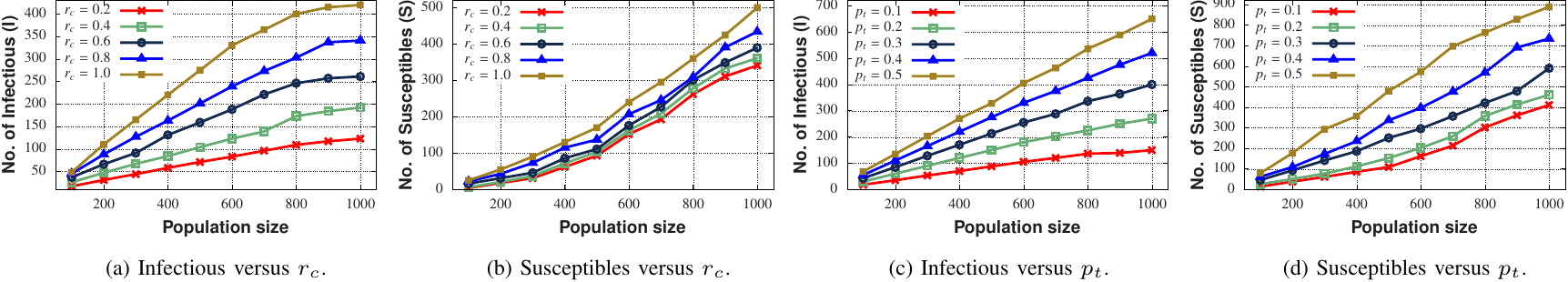} 
\caption{Impact of rate of contact ($r_c$) and probability of transmission ($p_t$) on number of infectious and susceptibles.} 
\label{rate_contact}
\end{figure*}

\begin{figure*}[ht]
\centering
\includegraphics[scale=1]{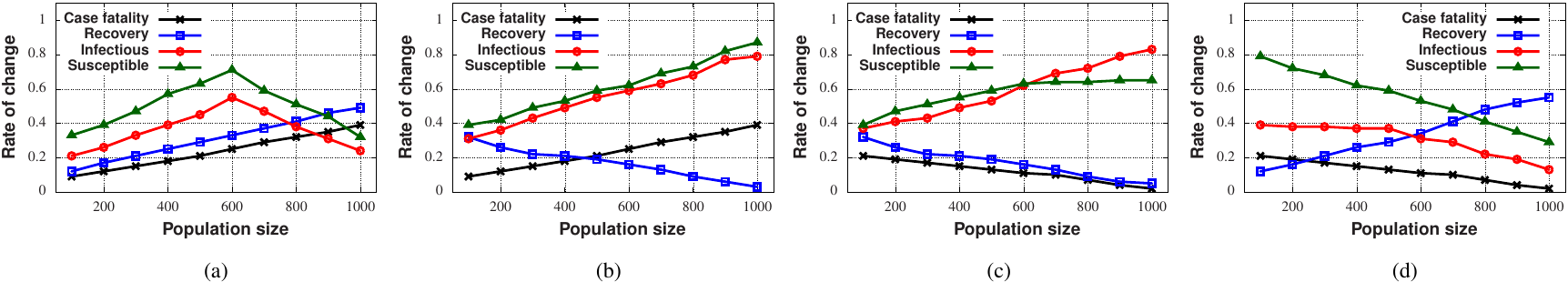}  
\caption{Impact of case fatality rate and recovery rate on infectious and susceptible. (a) having both case fatality and recovery increasing, (b) increasing case fatality and decreasing recovery, (c) increasing recovery and decreasing case fatality, and (d) both decreasing.} 
\label{case_fatality}
\end{figure*}

\subsubsection{Impact of rate of contact and probability of transmission}
The rate of contact and the probability of transmission play a vital role in determining the infection spread of an epidemic like COVID-19 in a particular area. Therefore, this paper calculates the number of infected and susceptible at a varying rate of contact and  probability of transmission separately. Firstly, we put the probability of transmission as a constant $(p_t=0.3)$ and the rate of contact as a variable. Part (a) of Fig.~\ref{rate_contact} illustrates that the increment in the infectious is nearly linear with the rate of contact and population size. At higher population size beyond $800$, the curve starts flattening as the infection has covered roughly half of the population. Similarly, a linear increment is observed for susceptible to the increase in the rate of contact.

Further, parts (c) and (d) of Fig.~\ref{rate_contact} show that the transmission probability is highly sensitive in determining the rate of infection spread and increment in the susceptible. We have considered the rate of contact ($r_c=1$) for calculating the value of infectious and susceptible. Part (c) of Fig.~\ref{rate_contact} demonstrates that the infectious count is directly proportional to the transmission probability and shows an exact linear increment with the increase in the population size. In a similar pattern, part (d) of Fig.~\ref{rate_contact} illustrates that the number of susceptible increases with the increase in the transmission probability as the chance to get infected is higher when the population increases. 

\subsubsection{Impact of case fatality rate and recovery rate}
Next, we evaluate the impact of the case fatality rate and recovery rate on the rate of change in infectious and susceptible. The case fatality rate and recovery rate forms following four cases
\begin{enumerate}
 \item \textbf{Both case fatality rate and recovery rate increase:} When both case fatality rate and recovery rate increase, then the rate of infectious and susceptible people are increased simultaneously. The susceptible population changes more rapidly to that of infectious. The increment in both infectious and susceptible continues up till a thresholding limit, where the influence of the case fatality rate start degrading to that of recovery rate. Part (a) of Fig.~\ref{case_fatality} shows the variation in infectious and susceptible when both case fatality and recovery rates increases. 
 \item \textbf{Case fatality rate increase and recovery rate decrease:} It is the most adverse situation for an epidemic breakout, where the case fatality is increasing, and recovery is decreasing. This clearly indicates that the epidemic is in the worst phase, and it can infect the entire population in a particular area. Part (b) of Fig.~\ref{case_fatality} illustrates the adverse situation where both infectious and susceptible are increasing.
 \item \textbf{Both case fatality rate and recovery rate decrease:} This is similar to that of the previously discussed case, but one positive thing that mitigates form severe damage is the decrement in the case fatality rate. Part (c) of Fig~\ref{case_fatality} indicates that the simultaneous decrement in both case fatality rate and recovery rate increases the infectious and susceptible. 
 \item \textbf{Case fatality rate decreases and recovery rate increases:} The last combination for case fatality and recovery rate is the case when the case fatality rate decreases and recovery rate increases. It is the most favourable situation where there is a clear indication that the epidemic is reducing. Part (d) of Fig.~\ref{case_fatality} illustrates that both infectious and susceptibles decrease under such circumstances.
\end{enumerate}


\section{Conclusion}\label{con1}
The pandemic COVID-19 is spreading at a higher pace around the globe. The empirical and data analysis in this paper tries to figure out the impact of various parameters such as the probability of transmission, rate of contacts, infectious and susceptible. The results indicate that the spread of COVID-19 can be analysed, modelled, and represented in the tabular and graphical format. The data analysis provides the complete in-site of the coronavirus spread and also highlights the reproduction number of COVID-19 is between $1.5$ to $4$, and its fatality rate is low. The paper covers those models that are widely in the various estimations in an epidemic and identifies the advantage and disadvantage of one model over others. The representation provides a more natural way to visualise the actual effect of the epidemic spread.

\section*{Acknowledgements}
The authors thank to Mr. Vinit Kumar Patra for doing the implementation and data preprocessing.

\bibliographystyle{IEEEtran}
\bibliography{final}
\vspace{-1.5cm}
\begin{IEEEbiography}
[{\includegraphics[height=.8in,keepaspectratio]{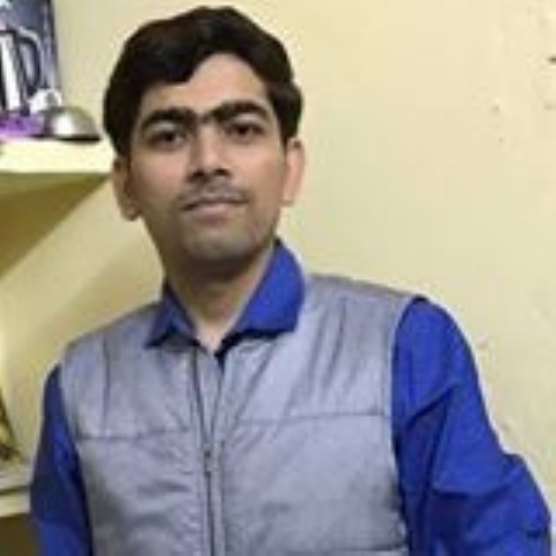}}] 
{~\\Rahul Mishra} received M.Tech. degree in Computer Science and Engineering from Madan Mohan Malaviya Univesity of Technology, U.P., India, in 2017. Presently, he is pursuing Ph.D. in Computer Science and Engineering, IIT (BHU) Varanasi, India. His research interests include wireless sensor networks, deep learning, and fog computing.
\end{IEEEbiography}
\vspace{-2.0cm}
\begin{IEEEbiography}
[{\includegraphics[height=.8in,keepaspectratio]{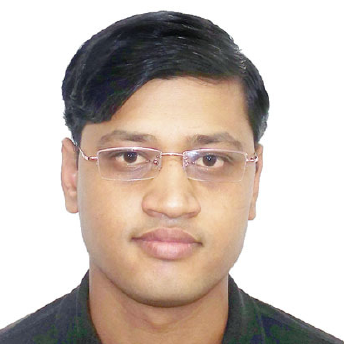}}]
{~\\Hari Prabhat Gupta} received  Ph.D degree in Computer Science and Engineering from IITG, Guwahati, India. He is currently working as Assistant Professor in Department of Computer Science and Engineering, IIT (BHU), Varanasi, India. His research interests include wireless sensor networks, machine learning, and distributed algorithms.
\end{IEEEbiography}
\vspace{-2.0cm}

\begin{IEEEbiography} 
[{\includegraphics[height=.8in,clip,keepaspectratio]{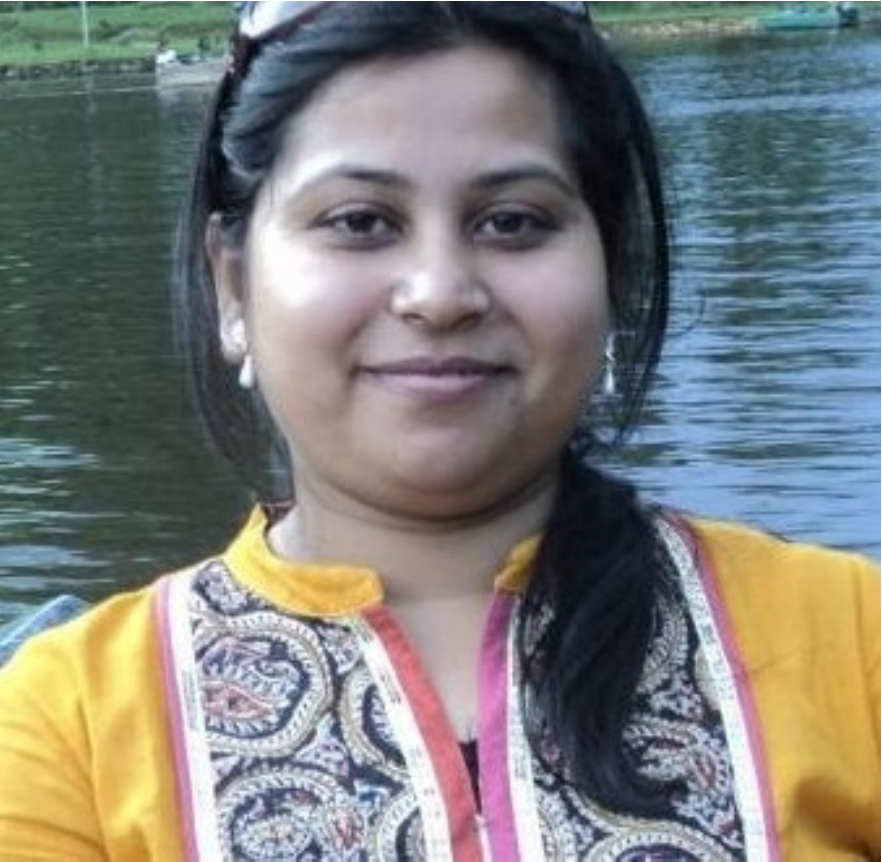}}]
{~\\Tanima Dutta} received  Ph.D degree in Computer Science and Engineering from IITG, Guwahati, India. She is currently working as Assistant Professor in Department of Computer Science and Engineering, IIT (BHU), Varanasi, India. Her research interests include multimedia, digital forensics, wireless sensor networks, and algorithms.
\end{IEEEbiography}
\end{document}